\documentclass[a4paper, preprint, preprintnumbers, superscriptaddress, showpacs, showkeys, nofootinbib, floatfix, twocolumn, 10pt]{revtex4-1}
\usepackage[utf8]{inputenc}
\usepackage[english]{babel}
\usepackage[dvips]{graphicx}
\usepackage{xcolor,url,amsthm,amssymb,amsmath,mathtools,latexsym,appendix,physics,bbm,enumerate}

\renewcommand{\exp}{\vb{e}}

\allowdisplaybreaks

\begin{document}

\title[LAQC for X states]{Local available quantum correlations of non-symmetric X states}

\author{David M. Bellorin R.}
\affiliation{Departamento de F\'{\i}sica, Universidad Sim\'on Bol\'{\i}var, AP 89000, Caracas 1080, Venezuela.}

\author{Hermann L. Albrecht Q.}
\email[Corresponding author: ]{albrecht@usb.ve}
\affiliation{Departamento de F\'{\i}sica, Universidad Sim\'on Bol\'{\i}var, AP 89000, Caracas 1080, Venezuela.}

\author{Douglas F. Mundarain}
%\email{dmundarain@ucn.cl}
\affiliation{Departamento de F\'{\i}sica, Universidad Cat\'olica del Norte, Casilla 1280, Antofagasta, Chile.}

\date{June 10, 2022}% It is always \today, today,

\begin{abstract}
Local available quantum correlations (LAQC), as defined by Mundarain et al., are analyzed for non-symmetric 2-qubit X  states, that is, X-states that are not invariant under the exchange of subsystems and therefore have local Bloch vectors whose norms are different. A simple analytic expression for their LAQC quantifier is obtained. As an example, we analyze the local application of the amplitude damping channel for Werner states and general X states. Although this local quantum channel can create quantum discord in some cases, no such outcome is possible for LAQC, which hints toward their monotonicity under LOCC operations. This work, along with our previous result for so-called symmetric and anti-symmetric X states, completes the pursuit of exact analytical expressions for the LAQC quantifier for 2-qubit X states.
\end{abstract}
\preprint{SB/F/495-22%, arXiv:2204.07552
}
\maketitle

%%%%%%%%%%%%%%%%%%%%%%%%%%%%%%%%%%%%%%%%%%%%%%
%%%%%%%%%%%%%%%%%%%%%%%%%%%%%%%%%%%%%%%%%%%%%%
%%%%%%%%%%%%% Introducción %%%%%%%%%%%%%%%%%%%
%%%%%%%%%%%%%%%%%%%%%%%%%%%%%%%%%%%%%%%%%%%%%%
%%%%%%%%%%%%%%%%%%%%%%%%%%%%%%%%%%%%%%%%%%%%%%

\section{Introduction}

Quantum information theory has been one of the most active physics research fields in the past couple of decades. A key ingredient for the advantages rendered by quantum mechanics within information theory is correlations. From the early days of the quantum revolution until the 2000s, only entanglement \cite{Horodecki-Ent} was the known quantum correlation. Since the development in 2001 of quantum discord \cite{qDiscord-Olliver, qDiscord-Henderson}, there has been an effort to study and develop new types of quantum correlations that can be applied in quantum information processing and communications \cite{Modi-qDiscord}.

Within these studies, the role of local measurements is crucial for developing new criteria for quanticity. A quantum correlation has to quantify the ability of a local observer to infer the results of another one from his own. The abovementioned quantum discord, for instance, is defined by comparing the quantum mutual information of the original state with a corresponding post-measurement state in the absence of readout. That is,  
\begin{equation}\label{eq:QD-Def}
D_A(\rho_{AB})\equiv \min_{\qty{\Pi_i^A}}\qty{I(\rho_{AB})-I[(\Pi^A\otimes\mathbbm{1}_2)\rho_{AB}]},
\end{equation}
where
\begin{equation}\label{eq:InfoMutua}
I(\rho_{AB})\equiv S\qty(\rho_A)+S\qty(\rho_B)-S\qty(\rho_{AB}),
\end{equation}
with $\rho_A$ and $\rho_B$ the reduced operators, i.e. marginals. 

In the above definition \eqref{eq:QD-Def}, the local measurement is performed on subsystem A, and the corresponding post-measurement states is usually referred to as classical-quantum (or A-classical) state,
\begin{equation}\label{eq:Estados_Cl-Q}
\rho^{cq}_{AB}=\sum_i\, p_i\,\dyad{i}\otimes\rho_B^i= \sum_i\, p_i\,\Pi^{(A)}_i\otimes\rho_B^i,
\end{equation}
\noindent{}and all such 2-qubit states constitute a set, denoted by $\Omega_o$. 

Analogously, quantum discord can also be established for a local measurement on subsystem B:
\begin{equation}\label{eq:QD-Def-B}
D_B(\rho_{AB})\equiv \min_{\{\Pi_i^B\}}\left\{I(\rho_{AB})-I[(\mathbbm{1}_2\otimes\Pi^B)\rho_{AB}]\right\},
\end{equation}
which leads to defining the set $\Omega_o'$ for quantum-classical (B-classical) states as in \eqref{eq:Estados_Cl-Q}. There have been other quantum correlations defined using the set $\Omega_o$ (or  $\Omega_o'$) \cite{GeomDisc,GD_TraceDist, GD_BuresDist} called Geometric Discords. Since the comparison involves local measurements on only one of the subsystems, they are in general not symmetric, $D_A(\rho_{AB})\neq{}D_B(\rho_{AB})$.

Trying to overcome this asymmetry, another set of quantum correlations and quantifiers was developed \cite{Luo_MID, Wu_AMID, Girolami_AMID, Wu_ComplementaryBases} by considering post-measurement states in the absence of readout whose measurements were performed locally on both subsystems:
\begin{equation}\label{eq:Estados-Clasicos}
\rho_{AB}^c = \sum_{i,j} p_{ij} \Pi^{(A)}_i\otimes\Pi^{(B)}_j.
\end{equation}
\noindent{}Such states are called classical states. A special subset is defined when $p_{ij}$ can be factorized so that
\begin{equation}
    \rho_{AB}^{\Pi}=\sum_{i,j} p_{i} \Pi^{(A)}_i\otimes{}p_{j}\Pi^{(B)}_j=\rho^{(A)}\otimes\rho^{(B)}.
\end{equation}
\noindent{}Such states are referred to as product or uncorrelated states, and the sets of classical and product 2-qubit states are labeled as $\Omega_c$ and $\Omega_{p}$, respectively.

In \cite{LAQC}, Mundarain and Ladrón de Guevara developed the so-called Local Available Quantum Correlations (LAQC). It is  a symmetric quantum correlation slightly different from the one presented by Wu et al. in \cite{Wu_ComplementaryBases}, defined in terms of mutual information of local bipartite measurements on the complementary basis of a previously determined optimal computational one.

This work is focused on an important family of 2-qubit states, the so-called X states \cite{EstadosX}
\begin{eqnarray}\label{eq:estadosX}
\rho^X =\mqty(
   \rho_{11} & 0 & 0 & \rho_{14} \\
   0 & \rho_{22} & \rho_{23} & 0 \\
   0 & \rho_{32} & \rho_{33}& 0 \\
   \rho_{41}& 0 & 0 & \rho_{44}).
\end{eqnarray}
These seven-parameter 2-qubit states have been extensively studied and used in QIT \cite{Quesada-XStates}. One of the main reasons for this is that any arbitrary 2-qubit state $\rho_{AB}$ can be mapped to a $\rho^X$ while preserving its main characteristics, e.g. quantum correlations \cite{XStates-Entanglement, Hedemann-XStates}.

Moreover, the calculation of several quantum correlation quantifiers is simpler for X states. For instance, concurrence was introduced by Wootters \cite{Wooters_Concurrence} as a \emph{bona fide} entanglement measure defined as
\begin{equation}\label{eq:DefConcurrencia}
\mathcal{C}\qty(\rho_{AB}) \equiv \max\qty{0,\lambda_1-\lambda_2-\lambda_3-\lambda_4},
\end{equation}
\noindent{}where $\qty{\lambda_i}$ are the decreasing ordered eigenvalues of
\begin{equation}
    \mathbf{R}=\sqrt{\sqrt{\rho_{AB}}\,\tilde{\rho_{AB}}\sqrt{\rho_{AB}}},
\end{equation}
with $\tilde{\rho_{AB}} = (\sigma_y\otimes\sigma_y)\rho_{AB}^*(\sigma_y\otimes\sigma_y)$, $\rho_{AB}^*$ the complex conjugate of $\rho_{AB}$, and $\sigma_y$ the corresponding Pauli matrix \eqref{eq:MatricesPauli}. A direct calculation shows that this entanglement measure takes a much simpler expression for X states \cite{EstadosX}:
\begin{equation}\label{eq:Concurrencia_X}
\mathcal{C}_X = \frac{1}{2}\,\max \left\{0,\mathcal{C}_1,\mathcal{C}_2\right\}
\end{equation}
\noindent{where}
\begin{equation*}
\mathcal{C}_1\equiv 2\left(|\rho_{14}|-\sqrt{\rho_{22}\,\rho_{33}}\right)\qc \mathcal{C}_2\equiv 2\left(|\rho_{23}|-\sqrt{\rho_{11}\,\rho_{44}}\right).
\end{equation*}

Regarding quantum discord, a closed analytical expression for a generic 2-qubit state cannot be obtained \cite{GirolamiAdesso-QD-Xstates}. For X states, although so far there is no such result, several approximations have been proposed \cite{Ali-QD-Xstates, Quesada-XStates, Lu-QD-Xstates-Ex2, Rau-Vinjanampathy_QD_2012, Li-QD-Xstates, Huang-QD-Xstates-WorstCaseScenario, Revisiting_QD, MaldonadoTrapp-QD, Rau_QD_2018}.  In this article, we use the
approximation introduced by Quesada et al. \cite{Quesada-XStates}, which we briefly present in an appendix, to determine the quantum discord and compare it with our results for the LAQC quantifier.

Moreover, since X states exhibit a $\mathfrak{su}(2)\times\mathfrak{su}(2)\times\mathfrak{u}(1)$ invariance symmetry \cite{Rau_2009-Xstates_algebra}, the definition of an X state can be generalized beyond \eqref{eq:estadosX}. Also, this algebraic characterization allows for defining general local and global quantum channels that map an X-state into another one. 

We can split  the family of X states using different criteria. From a geometrical perspective, Kelleher et al. \cite{Kelleher_Xstates-GeomPersp} introduced a classification of X states using the so-called perp-sets, which are a particular type of geometric hyperplanes of the symplectic
polar space of order two and rank two. On the other hand, we propose dividing them by considering whether their local Bloch vectors have equal norms. 

This criterion defines two sets, with the one with equally-normed local Bloch vectors exhibiting symmetry under subsystems exchange $\vb{A}\leftrightarrow\vb{B}$. For such states, we have that $\rho^{X}_{22}=\rho^{X}_{33}$ or $\rho^{X}_{11}=\rho^{X}_{44}$, depending on whether their local Bloch vectors are parallel or anti-parallel. Hence, we labeled them as symmetric and anti-symmetric X-states, respectively. In a previous paper \cite{LAQC_Xstates-sym}, we determined the LAQC's quantifier for this type of X state. Our goal here is to focus on the remaining set. That is those X states whose local Bloch vectors have different magnitude. Since they no longer exhibit the abovementioned symmetry, we label this set as non-symmetric X states.

We have structured this paper as follows. We start reviewing the procedure for determining the local available quantum correlations quantifier \cite{LAQC}. Next, we study its calculation for X states whose local Bloch vectors have a different norm. Then, we present some examples and discuss the effect of local quantum channels on the LAQC’s quantifier. Finally, we summarize the conclusions from our results. We have included an appendix briefly introducing the approximation for quantum discord of X states that Quesada et al. \cite{Quesada-XStates} proposed and which we use for computing it when comparing it to LAQC.

%%%%%%%%%%%%%%%%%%%%%%%%%%%%%%%%%%%%%%%%%%%%%%
%%%%%%%%%%%%%%%%%%%%%%%%%%%%%%%%%%%%%%%%%%%%%%
%%%%%%%%% Intro LAQC 2-qubits %%%%%%%%%%%%%%%%
%%%%%%%%%%%%%%%%%%%%%%%%%%%%%%%%%%%%%%%%%%%%%%
%%%%%%%%%%%%%%%%%%%%%%%%%%%%%%%%%%%%%%%%%%%%%%
\section{\label{sec:LAQCs-2qubits}Local available quantum correlations of 2-qubits}

Any density operator $\rho$ can always be written in terms of different local bases. For bipartite qubit systems, given two general bases, $\qty{\ket{k,m}}$ and $\qty{\ket{\mu,\nu}}$, it's density operator $\rho_{AB}$ can be written in either basis as
\begin{equation}\label{eq:rho_Bases}
  \rho_{AB}=\sum_{klmn}\rho_{kl}^{mn}\,\dyad{k,m}{l,n}=\sum_{\mu\nu\eta\gamma}R_{\mu\eta}^{\nu\gamma}\,\dyad{\mu,\nu}{ \eta,\gamma},
\end{equation}
where $k, l, m, n,\mu,\nu,\eta,\gamma \in \{0,1\}$. Both bases are equivalent under a local unitary transformations
\begin{equation}\label{eq:BasesTransform}
    \ket{\mu,\nu}=\qty(\mathbbm{U}_{\vb{A}}\otimes{U}_{\vb{B}})\ket{k,m}\qc\mathbbm{U}_A,\mathbbm{U}_B\in{U}(2).
\end{equation}
Any such basis can be used as a computational one, whose elements are the eigenvectors of $\sigma_{\vu{u}}\equiv\va*{\sigma}\cdot\vu{u}$, with $\va*{\sigma}$ the vector whose local components are the Pauli matrices
\begin{equation}\label{eq:MatricesPauli}
    \sigma_1=\mqty(\pmat{1})\qc\sigma_2=\mqty(\pmat{2})\qc\sigma_3=\mqty(\pmat{3}),
\end{equation}
and $\vu{u}\in\mathbbm{E}^3$ is a generic unitary vector, whose components can be written in terms of the parametrization of the respective $U(2)$ transformation.

For classical states, there exists a local basis for which the density operator $\rho_{AB}^c$ \eqref{eq:Estados-Clasicos} is diagonal. One can define a particular classical state $X_\rho\in\Omega_c$ related to $\rho_{AB}$ as the one induced by a measurement $\mathbbm{M}_o$ that minimizes
\begin{equation}\label{eq:S(rho||X)}
S\qty(\rho_{AB}||X_\rho)=\min_{\Omega_c}S\qty(\rho_{AB}||\chi_\rho),
\end{equation}
\noindent{}where $S\qty(\rho||\chi)=-\mathrm{Tr}(\rho\mathrm{log}_2\chi)-S(\rho)$ is the relative entropy and
\begin{subequations}\label{eq:Chi_rho}
\begin{align}
\chi_\rho =&\sum_{\mu\nu}R_{\mu\nu}\,\dyad{\mu,\nu)},\\ 
R_{\mu\nu} =& \expval{\rho_{AB}}{\mu,\nu)}\label{eq:Coeff_Rij}.
\end{align}
\end{subequations}
\noindent{}The minimization in \eqref{eq:S(rho||X)} is equivalent to determining the coefficients $\qty\big{R_{ij}^{opt}}$ when $\chi_\rho=X_\rho$. These coefficients are associated with a new basis, labeled as the optimal computational basis $\qty\big{\ket{i,j}^{opt}}$, and the local available quantum correlations (LAQC) are defined in terms of it.

To determine $\qty\big{\ket{i,j}^{opt}}$, we define a general orthonormal basis for each subsystem:
\begin{equation}\label{eq:BaseOrtonormalGen}
\begin{aligned}
\ket{\mu_0^{(n)}} &= \cos\frac{\theta_n}{2}\ket{0^{(n)}} +\sin\frac{\theta_n}{2}\exp^{i\phi_n}\ket{1^{(n)}},\\
\ket{\mu_1^{(n)}} &= -\sin\frac{\theta_n}{2}\ket{0^{(n)}} +\cos\frac{\theta_n}{2}\exp^{i\phi_n}\ket{1^{(n)}},
\end{aligned}
\end{equation}
\noindent{}where $n=1$ denotes subsystem \textbf{A} and $n=2$, subsystem \textbf{B}. Such basis is the result of applying
\begin{equation}\label{eq:U(2)-Gen}
    \mathbbm{U}_n = \mqty(\cos\frac{\theta_n}{2} & -\sin\frac{\theta_n}{2}  \\
   \sin\frac{\theta_n}{2}\exp^{i\phi_n} & \cos\frac{\theta_n}{2}\exp^{i\phi_n})\;\;\in\;U(2)
\end{equation}
to the original computational basis of each subsystem. The classical correlations quantifier is given by  \cite{LAQC, Modi-RelativeEntropy}
\begin{equation}\label{eq:CorrClasicas}
\mathcal{C}(\rho) = I(X_\rho).
\end{equation}
The mutual information \eqref{eq:InfoMutua} may be written as 
\begin{equation}\label{eq:Info_Mutua-Prob}
\begin{split}
I\qty(\rho_{AB}) =& \sum_{i,j} P_{\theta,\phi}(i_A,j_B)\\
& \qq{}\times\log_2\qty[\frac{P_{\theta,\phi}(i_A,j_B)}{P_{\theta_{A},\phi_{A}}(i_A)P_{\theta_B,\phi_B}(j_B)}],
\end{split}
\end{equation}
\noindent{}where 
\begin{equation}\label{eq:P_theta_phi}
    P_{\theta,\phi}(i_A,j_B) = \expval{\rho_{AB}}{\mu_i^{(1)},\mu_j^{(2)}}
\end{equation}
are the probability distributions corresponding to $\rho_{AB}$ and 
\begin{subequations}\label{eq:P_i_theta_phi}
\begin{align}
    P_{\theta_{A},\phi_{A}}(i_A) &= \expval{\rho_A}{\mu_i^{(1)}},\\
    P_{\theta_{B},\phi_{B}}(j_B) &= \expval{\rho_B}{\mu_i^{(2)}},
\end{align}
\end{subequations}
are the ones corresponding to its reduced operators $\rho_A$ and $\rho_B$. The minimization of the relative entropy \eqref{eq:S(rho||X)} required to define $X_{\rho}$ yields a minima for the classical correlations quantifier defined in \eqref{eq:CorrClasicas}. It is straightforward to realize that these probability distributions are directly related to the $\qty\big{R_{ij}^{opt}}$ coefficients when $\qty{\ket{\mu_i^{(1)},\mu_j^{(2)}}}$ is the optimal computational basis.

With the optimal computational basis determined, the state $\rho_{AB}$ is rewritten and the complementary basis defined as
\begin{eqnarray}\label{eq:u0-u1}
  \ket{u_0^{(m)}}&=&\frac{1}{\sqrt{2}}\qty(\ket{0^{(m)}}_{opt}+\exp^{i\Phi_m}\ket{1^{(m)}}_{opt}),\nonumber\\
  \ket{u_1^{(m)}}&=&\frac{1}{\sqrt{2}}\qty(\ket{0^{(m)}}_{opt}-\exp^{i\Phi_m}\ket{1^{(m)}}_{opt}).
\end{eqnarray}
This basis results from applying the transformation \eqref{eq:U(2)-Gen} with $\theta_n=\frac{\pi}{2}$ and a new angle $\phi_n\rightarrow\Phi_n$ via \eqref{eq:BasesTransform}. The corresponding probability distributions $P(i_A,j_B,\Phi)$ and marginal probability distributions $P^{(n)}(i_n,\Phi)$ analogue to \eqref{eq:P_theta_phi} and \eqref{eq:P_i_theta_phi} are determined. The maximization of $I\qty(\Phi_1,\Phi_2)$ \eqref{eq:Info_Mutua-Prob} corresponds to the LAQC quantifier:
\begin{equation}\label{eq:LAQC-quant}
    \mathcal{L}(\rho_{AB}) \equiv \max_{\qty{\Phi_1,\Phi_2}} I\qty(\Phi_1,\Phi_2).
\end{equation}

%%%%%%%%%%%%%%%%%%%%%%%%%%%%%%%%%%%%%%%%%%%%%%
%%%%%%%%%%%%%%%%%%%%%%%%%%%%%%%%%%%%%%%%%%%%%%
%%%%%%%%%%% LAQC non symm X states %%%%%%%%%%%
%%%%%%%%%%%%%%%%%%%%%%%%%%%%%%%%%%%%%%%%%%%%%%
%%%%%%%%%%%%%%%%%%%%%%%%%%%%%%%%%%%%%%%%%%%%%%
\section{LAQC of non-symmetric X states} \label{sec:LAQCs_X}

\vspace{5mm}
The 7-parameter 2-qubit X states can be mapped into a simpler 5-parameter subset as there are two phase parameters that can be removed via local transformations, as established by Zhou et al. \cite{Zhou-CanonicalXstates}. By defining
\begin{equation}
    \rho_{14}=w\,\exp^{i\xi_1}\qq{and} \rho_{23}=z\,\exp^{i\xi_2},
\end{equation}
along with labeling the diagonal elements as $a$, $b$, $c$, and $d$, respectively, the density matrix \eqref{eq:estadosX} is then given by
\begin{eqnarray}\label{eq:Estados_X-abcdzwxi1xi2}
    \rho^X =\mqty(
   a & 0 & 0 & w\,\exp^{i\xi_1} \\
   0 & b & z\,\exp^{i\xi_2} & 0 \\
   0 & z\,\exp^{-i\xi_2} & c& 0 \\
   w\,\exp^{-i\xi_1}& 0 & 0 & d),
\end{eqnarray}
where $a,b,c,w,z, \xi_1,\xi_2\,\in\,\mathbbm{R}$ are the seven independent real parameters labeling X states. $\xi_1$ and $\xi_2$ are two abovementioned phase parameters that can be removed via local transformations \cite{Zhou-CanonicalXstates}. Therefore, the density matrix of X states is written as
\begin{eqnarray}\label{eq:Estados_X-abcdzw}
    \rho^X =\mqty(
   a & 0 & 0 & w \\
   0 & b & z & 0 \\
   0 & z & c& 0 \\
   w& 0 & 0 & d),
\end{eqnarray}
where
\begin{subequations}
\begin{align}
    &a,b,c,d\geq0,\\
    &d=1-\qty(a+b+c),\\
    &z\leq\sqrt{bc}\qq{and}w\leq\sqrt{ad},
\end{align}
\end{subequations}
are the constraints on these parameters so $\rho^X$ is a well-behaved density matrix, i.e. is hermitian, has $\Tr\qty(\rho^X)=1$, and is semi-positive definite.

By introducing
\begin{equation}
    \sigma_0 =\mathbbm{1}_2 = \mqty(\imat{2}),
\end{equation}
a Hermitian base for the space of linear operators acting on the qubit Hilbert space is obtained, which allows to write any qubit state as
\begin{equation}\label{eq:Bloch-Qubit}
    \rho=\sum_{i}c_{i}\dyad{i}=\frac{1}{2}\qty(\mathbbm{1}_2+\va*{\tau}\cdot\va*{\sigma}),
\end{equation}
where $\va*{\tau}$ is the Bloch vector of $\rho$. This representation can be readily extended to 2-qubit systems, often referred to as the Fano form or Fano-Bloch representation \cite{Fano1983},
\begin{equation}\label{eq:2qubits-Bloch}
\begin{split}
\rho = \frac{1}{4}\Bigg[ \mathbbm{1}_4 + \qty(\va{x}\cdot\va*{\sigma})\otimes\mathbbm{1}_2& + \mathbbm{1}_2\otimes \qty(\va{y}\cdot\va*{\sigma})\\ &+\va*{\sigma}\cdot\mathbbm{T}\cdot\va*{\sigma})\Bigg],
\end{split}
\end{equation}
where $\va{x}$ and $\va{y}$ are the local Bloch vectors of subsystems \textbf{A} and \textbf{B}, respectively, and $\mathbbm{T}$ is the correlations tensor. The components of the local Bloch vectors $x_i$ and $y_i$ as well as the ones of the correlations tensor $T_{ij}$ are given by
\begin{subequations}\label{eq:ParametrosBloch-2qubits}
\begin{align}
    x_i&=\Tr\qty[\rho\qty(\sigma_i\otimes\mathbbm{1}_2)],\\
    y_i&=\Tr\qty[\rho\qty(\mathbbm{1}_2\otimes\sigma_i)],\\
    T_{ij}&=\Tr\qty[\rho\qty(\sigma_i\otimes\sigma_j)].
\end{align}
\end{subequations}

For X states \eqref{eq:Estados_X-abcdzw}, we have that
\begin{equation}\label{eq:estadosX-Bloch}
\begin{split}
\rho^X = \frac{1}{4}\Bigg( \mathbbm{1}_4 + x_3\,\sigma_3\otimes\mathbbm{1}_2& + \mathbbm{1}_2\otimes y_3\,\sigma_3\\ &+\sum_{n=1}^3 T_{n}\sigma_n\otimes\sigma_n\Bigg).
\end{split}
\end{equation}
%where as for the 7-parameter ones \eqref{eq:Estados_X-abcdzwxi1xi2} two extra terms need to be included, namely
%\begin{equation}
%    T_{12}\sigma_1\otimes\sigma_2\qq{and}T_{21}\sigma_2\otimes\sigma_1.
%\end{equation}
%From these expressions, it can be inferred after some thought that the local transformations involved in mapping a 7-parameter X state into a 5-parameter one are those that diagonalize the correlations tensor, $\mathbbm{T}$. 
%
It is an alternative parametrization to \eqref{eq:Estados_X-abcdzw}, with $x_3$, $y_3$, and $T_n$ related to the previous $a,b,c,d,z,$ and $w$ by
\begin{equation}\label{eq:Bloch_Xstates-abcdwz}
\begin{aligned}
    x_3&=a+b-c-d,\\
    y_3&=a-b+c-d,
\end{aligned}\qq{}
\begin{aligned}
    T_1&=2(z+w),\\
    T_2&=2(z-w),\\
    T_3&=a-b-c+d.
\end{aligned}
\end{equation}

As was previously mentioned, 2-qubit X-states can be divided in two categories, depending on whether their local Bloch vectors have equal norm or not. The symmetrical and anti-symmetrical category were already analyzed in \cite{LAQC_Xstates-sym}. We will focus our attention on X states with  $\abs{x_3}\neq{}\abs{y_3}$, which is equivalent to requiring that $a\neq{}d$ and $b\neq{}c$. We label this set as non-symmetric X states, $\rho^{X_{ns}}$. Since such states are not invariant under subsystem exchange $\vb{A}\leftrightarrow\vb{B}$, all four parameters $\theta_1$, $\theta_2$, $\phi_1$, and $\phi_2$ in \eqref{eq:BaseOrtonormalGen} have to be taken into account. A direct computation of the $R_{ij}$ coefficients \eqref{eq:Chi_rho} for the general computational basis \eqref{eq:BaseOrtonormalGen} for $\rho^X$ \eqref{eq:Estados_X-abcdzw} leads to
\begin{subequations}
\begin{align}
     R_{00}%=\,&\frac{1}{4}\qty(1+\cos\theta_1)\qty[\qty(a-b-c+d)\qty(1+\cos\theta_2)+b-d]\nonumber
%        \\
%    &+\frac{1}{2}\qty[w\cos\qty(\phi_1+\phi_2)+z\cos\qty(\phi_1-\phi_2)]\sin\theta_1\sin\theta_2\nonumber
%        \\
%    &+\frac{1}{2}(c-d)\qty(1+\cos\theta_2)+d\nonumber
%        \\
        =\,& \frac{1}{4}\Big[1+x_3\cos\theta_1 +y_3\cos\theta_2+T_3\cos\theta_1\cos\theta_2\nonumber
        \\ 
        &\qq{}+\sin\theta_1\sin\theta_2\big(T_1\cos\phi_1\cos\phi_2\nonumber
        \\ 
        &\qq{\hspace{10mm}}+T_2\sin\phi_1\sin\phi_2\big)\Big],
\\
    R_{01}%=\,&-\frac{1}{4}\qty(1+\cos\theta_1)\qty[\qty(a-b-c+d)\qty(1+\cos\theta_2)+a-c]\nonumber
%        \\
 %   &-\frac{1}{2}\qty[w\cos\qty(\phi_1+\phi_2)+z\cos\qty(\phi_1-\phi_2)]\sin\theta_1\sin\theta_2\nonumber
%        \\
%    &-\frac{1}{2}(c-d)\qty(1+\cos\theta_2)+c\nonumber
%        \\
        =\,& \frac{1}{4}\Big[1+x_3\cos\theta_1 -y_3\cos\theta_2-T_3\cos\theta_1\cos\theta_2 \nonumber
        \\ 
        &\qq{}-\sin\theta_1\sin\theta_2\big(T_1\cos\phi_1\cos\phi_2\nonumber
        \\ 
        &\qq{\hspace{10mm}}+T_2\sin\phi_1\sin\phi_2\big)\Big],
\\
    R_{10}%=\,&-\frac{1}{4}\qty(1+\cos\theta_1)\qty[\qty(a-b-c+d)\qty(1+\cos\theta_2)-b+d]\nonumber
%        \\
%    &-\frac{1}{2}\qty[w\cos\qty(\phi_1+\phi_2)+z\cos\qty(\phi_1-\phi_2)]\sin\theta_1\sin\theta_2\nonumber
%        \\
%    &-\frac{1}{2}(c-d)\qty(1+\cos\theta_2)+b\nonumber
%        \\
        =\,& \frac{1}{4}\Big[1-x_3\cos\theta_1 +y_3\cos\theta_2-T_3\cos\theta_1\cos\theta_2 \nonumber
        \\ 
        &\qq{}-\sin\theta_1\sin\theta_2\big(T_1\cos\phi_1\cos\phi_2\nonumber
        \\ 
        &\qq{\hspace{10mm}}+T_2\sin\phi_1\sin\phi_2\big)\Big],
\\
    R_{11}%=\,&\frac{1}{4}\qty(1+\cos\theta_1)\qty[\qty(a-b-c+d)\qty(1+\cos\theta_2)-a+c]\nonumber
%        \\
%    &+\frac{1}{2}\qty[w\cos\qty(\phi_1+\phi_2)+z\cos\qty(\phi_1-\phi_2)]\sin\theta_1\sin\theta_2\nonumber
%    \\
%    &-\frac{1}{2}(a-b)\qty(1+\cos\theta_2)+a\nonumber
%        \\
        =\,& \frac{1}{4}\Big[1-x_3\cos\theta_1 -y_3\cos\theta_2+T_3\cos\theta_1\cos\theta_2\nonumber
        \\ 
        &\qq{}+\sin\theta_1\sin\theta_2\big(T_1\cos\phi_1\cos\phi_2\nonumber
        \\ 
        &\qq{\hspace{10mm}}+T_2\sin\phi_1\sin\phi_2\big)\Big].
\end{align}
\end{subequations}
For the reduced matrices we have that
\begin{subequations}
\begin{align}
    R_{\smqty{0\\1}}^{(\vb{A})}&=\frac{1}{2}\qty(1\pm{}x_3\cos\theta_1),
\\
    R_{\smqty{0\\1}}^{(\vb{B})}&=\frac{1}{2}\qty(1\pm{}y_3\cos\theta_2).
\end{align}
\end{subequations}

It is easily verified that for
\begin{eqnarray}\label{eq:Theta-OptimalCompBasis}
    \theta_i =(2k+1)\frac{\pi}{2} \qq{and} \theta_j=\tilde{k}\pi,
\end{eqnarray}
with $k=0,1$, $\tilde{k}=0,1,2$, and $i,j=1,2$, %so that
%\begin{equation}
%     \theta_2 - \theta_1= \pm(2\kappa+1)\frac{\pi}{2}.
%\end{equation}
%the argument of each logarithm involved in \eqref{eq:Info_Mutua-Prob} is given by
we have that
\begin{equation}
    \frac{R_{ij}}{R_{i}^{(\vb{A})}R_{j}^{(\vb{B})}}=1.
\end{equation}
Therefore, by choosing $\theta_1$ and $\theta_2$ so that \eqref{eq:Theta-OptimalCompBasis} is satisfied, with $\phi_1$ and $\phi_2$ arbitrary, we obtain that the classical correlations quantifier $\mathcal{C}\qty(\rho^X)$ \eqref{eq:CorrClasicas} is
\begin{equation}\label{eq:CorrClass-Xstates-Asymmetric}
    \mathcal{C}\qty(\rho^{X_{ns}}) = 0.
\end{equation}
Since such angles correspond to possible optimal computational bases, we have to rewrite $\rho^{X_{ns}}$ \eqref{eq:estadosX-Bloch} in terms of them and then determine the corresponding probability distributions $P_{\theta_1,\theta_2}(i_A,j_B,\phi_1,\phi_2,\Phi_1,\Phi_2)$. There are two distinct possibilities for $\theta_1$ and $\theta_2$ for the computational basis,
\begin{equation}\label{eq:thetas-cases}
    \begin{cases}
    \theta_1&=0\qc \theta_2=\frac{\pi}{2},\\
    \theta_1&=\frac{\pi}{2}\qc \theta_2=0.
    \end{cases}
\end{equation}
All $\theta_i =(2k+1)\frac{\pi}{2}$ and $\theta_j=\hat{k}\pi$ lead to the same expressions of $P(i_A,j_B,\Phi_1,\Phi_2)$ for one of the above cases. %Therefore, we only present the density matrices for  $\theta_1=0$ and $\theta_2=\frac{\pi}{2}$, as well as for $\theta_1=\frac{\pi}{2}$ and $\theta_2=0$.
%
%\begin{widetext}
%\begin{subequations}
%\begin{align}
%    \rho^{X\,(opt)}_{\qty(0,\frac{\pi}{2})} =\,&\frac{1}{2}
%    \mqty(
%    a+b &(a-b)\exp^{-i\phi_2} & -(w+z)\exp^{-i\phi_1} & (w-z)\exp^{-i\qty(\phi_1+\phi_2)}\\
    %%%
%    (a-b)\exp^{i\phi_2} & a+b &  -(w-z)\exp^{-i\qty(\phi_1-\phi_2)} & (w+z)\exp^{-i\phi_1}\\
    %%%
%    -(w+z)\exp^{i\phi_1} &  -(w-z)\exp^{i\qty(\phi_1-\phi_2)}& c+d & (c-d)\exp^{-i\phi_2}\\
    %%%%
%    (w-z)\exp^{i\qty(\phi_1+\phi_2)} & (w+z)\exp^{i\phi_1} & (c-d)\exp^{i\phi_2} & c+d
%    ),\label{eq:rhoX-opt-theta1=0theta2=pi2}\\
%    \rho^{X\,(opt)}_{\qty(\frac{\pi}{2},0)} =\,& \frac{1}{2}
%     \mqty(
%    a+c &-(w+z)\exp^{-i\phi_2} & (a-c)\exp^{-i\phi_1} & (w-z)\exp^{-i\qty(\phi_1+\phi_2)}\\
    %%%
%    -(w+z)\exp^{i\phi_2} & b+d & -(w-z)\exp^{-i\qty(\phi_1-\phi_2)}  & (b-d)\exp^{-i\phi_1}\\
    %%%
%    (a-c)\exp^{i\phi_1} & -(w-z)\exp^{i\qty(\phi_1-\phi_2)} & a+c & (w+z)\exp^{-i\phi_2}\\
    %%%
%    (w-z)\exp^{i\qty(\phi_1+\phi_2)} & (b-d)\exp^{i\phi_1} & (w+z)\exp^{i\phi_2} & b+d
%    ).\label{eq:rhoX-opt-theta1=pi2theta2=0}
%\end{align}
%\end{subequations}
%\end{widetext}
For $\theta_1=\tilde{k}\pi$ and $\theta_2 =(2k+1)\frac{\pi}{2}$ $(k=0,1, \tilde{k}=0,1,2)$, we have that
\begin{subequations}
\begin{align}
    P_{\qty(0,\frac{\pi}{2})}(0,0,\Psi_1,\Psi_2) &= \frac{1}{4}\big(1+y_3\cos\Psi_2\nonumber\\
    &\qq{ }+T_2\sin\Psi_1\sin\Psi_2\big),
    \\
    P_{\qty(0,\frac{\pi}{2})}(0,1,\Psi_1,\Psi_2) &= \frac{1}{4}\big(1-y_3\cos\Psi_2\nonumber\\
    &\qq{ }-T_2\sin\Psi_1\sin\Psi_2\big),
    \\
    P_{\qty(0,\frac{\pi}{2})}(1,0,\Psi_1,\Psi_2) &= \frac{1}{4}\big(1+y_3\cos\Psi_2\nonumber\\
    &\qq{ }-T_2\sin\Psi_1\sin\Psi_2\big),
    \\
    P_{\qty(0,\frac{\pi}{2})}(1,1,\Psi_1,\Psi_2) &= \frac{1}{4}\big(1-y_3\cos\Psi_2\nonumber\\
    &\qq{ }+T_2\sin\Psi_1\sin\Psi_2\big),
\end{align}
\end{subequations}
where
\begin{eqnarray}
    \Psi_i\equiv\Phi_i-\phi_1\;\;\; (i=1,2)\qc \Psi_i\in\qty[-2\pi,2\pi].
\end{eqnarray} 
The corresponding marginal probabilities are given by
\begin{subequations}
\begin{align}
    P_{\qty(0,\frac{\pi}{2})}^{(\vb{A})}(0,\Psi_1,\Psi_2) &=\frac{1}{2} = P_{\qty(0,\frac{\pi}{2})}^{(\vb{A})}(1,\Psi_1,\Psi_2),\\
    P_{\qty(0,\frac{\pi}{2})}^{(\vb{B})}(0,\Psi_1,\Psi_2) &=\frac{1}{2}\qty(1+y_3\cos\Psi_2),\\ P_{\qty(0,\frac{\pi}{2})}^{(\vb{B})}(1,\Psi_1,\Psi_2)&=\frac{1}{2}\qty(1-y_3\cos\Psi_2).
\end{align}
\end{subequations}

Similar expressions are obtained when $\theta_1 =(2k+1)\frac{\pi}{2}$ and $\theta_2=\tilde{k}\pi$ $(k=0,1)$, with $x_3$ instead of the previous $y_3$.
\begin{subequations}
\begin{align}
    P_{\qty(\frac{\pi}{2},0)}(0,0,\Psi_1,\Psi_2) &= \frac{1}{4}\big(1+x_3\cos\Psi_1\nonumber\\
    &\qq{ }+T_2\sin\Psi_1\sin\Psi_2\big),
    \\
    P_{\qty(\frac{\pi}{2},0)}(0,1,\Psi_1,\Psi_2) &= \frac{1}{4}\big(1+x_3\cos\Psi_1\nonumber\\
    &\qq{ }-T_2\sin\Psi_1\sin\Psi_2\big),
    \\
    P_{\qty(\frac{\pi}{2},0)}(1,0,\Psi_1,\Psi_2) &= \frac{1}{4}\big(1-x_3\cos\Psi_1\nonumber\\
    &\qq{ }-T_2\sin\Psi_1\sin\Psi_2\big),
    \\
    P_{\qty(\frac{\pi}{2},0)}(1,1,\Psi_1,\Psi_2) &= \frac{1}{4}\big(1-x_3\cos\Psi_1\nonumber\\
    &\qq{ }+T_2\sin\Psi_1\sin\Psi_2\big),
\end{align}
\end{subequations}
and the corresponding marginal probabilities
\begin{subequations}
\begin{align}
    P_{\qty(\frac{\pi}{2},0)}^{(\vb{A})}(0,\Psi_1,\Psi_2) &=\frac{1}{2}\qty(1+x_3\cos\Psi_1),\\ P_{\qty(\frac{\pi}{2},0)}^{(\vb{A})}(1,\Psi_1,\Psi_2)&= \frac{1}{2}\qty(1-x_3\cos\Psi_1),\\
    P_{\qty(\frac{\pi}{2},0)}^{(\vb{B})}(0,\Psi_1,\Psi_2) &=\frac{1}{2}= P_{\qty(\frac{\pi}{2},0)}^{(\vb{B})}(1,\Psi_1,\Psi_2).
\end{align}
\end{subequations}

The maximization required in \eqref{eq:LAQC-quant} is readily achieved for $\Psi_i=\pm(2k+1)\frac{\pi}{2}$, with $i=1,2$, and $k=0,1$. Therefore, we  have a single expression for the LAQC quantifier of non-symmetric X states:
\begin{equation}\label{eq:LAQC-Asymm-Xstates}
\begin{aligned}
     \mathcal{L}\qty(\rho^{X_{ns}}) =\,& \frac{1+T_2}{2}\log_2\qty(1+T_2)\\ 
     &\qq{}+ \frac{1-T_2}{2}\log_2\qty(1-T_2).
\end{aligned}
\end{equation}
Notice that this expression corresponds to the function $g_2$ \eqref{eq:gi(Ti)} defined in \cite{LAQC_Xstates-sym}.

%%%%%%%%%%%%%%%%%%%%%%%%%%%%%%%%%%%%%%%%%%%%%%
%%%%%%%%%%%%%%%%%%%%%%%%%%%%%%%%%%%%%%%%%%%%%%
%%%%%%%%%%%%%%% Ejemplos %%%%%%%%%%%%%%%%%%%%%
%%%%%%%%%%%%%%%%%%%%%%%%%%%%%%%%%%%%%%%%%%%%%%
%%%%%%%%%%%%%%%%%%%%%%%%%%%%%%%%%%%%%%%%%%%%%%
\section{LAQC and Local Quantum Channels: Amplitude Damping}

The amplitude damping (AD) channel \cite{Nielsen-QIT} describes the process of energy dissipation (e. g. spontaneous emission) into the environment. The Kraus operators \cite{Kraus-Article} for this quantum operation on one qubit are given by
\begin{equation}\label{eq:Kraus_AD}
    \vb{E}^{(AD)}_0 = \mqty(1 & 0\\ 0 & \sqrt{1-p\,})\,\qc\quad \vb{E}^{(AD)}_1 = \mqty(0 & \sqrt{p\,}\\ 0 & 0),
\end{equation}
\noindent{}where $p$ is the channel's parameter, which can be thought of as the probability for the transition $\ket{1} \rightarrow \ket{0}$.

We are interested in applying the AD channel to a bipartite qubit system. Within the operator-sum representation, the extension is readily achieved via
\begin{equation}\label{eq:Kraus_Interaccion}
  \rho\longrightarrow \rho'=\sum_{i,j} \left(\vb{K}^{(A)}_i\otimes\vb{K}^{(B)}_j\right)\rho\left(\vb{K}^{(A)}_i\otimes\vb{K}^{(B)}_j\right)^\dagger,
\end{equation}
\noindent{}where $\vb{K}^{(I)}_i$ is the $i$-th Kraus operator of a given quantum channel acting the $I$ subsystem. Since we are interested in a local quantum operation, without loss of generality we focus in what follows on a quantum channel acting only on subsystem \textbf{B}. Therefore, we have that
\begin{equation}\label{eq:Kraus_AD-2qubit}
  \rho'=\sum_{i} \left(\mathbbm{1}\otimes\vb{E}_i\right)\rho\left(\mathbbm{1}\otimes\vb{E}_i\right)^\dagger.
\end{equation}

In \cite{Rau_2009-Xstates_algebra}, Rau established the algebraic characterization of 2-qubit X states in terms of the $\mathfrak{su}(2)\times{}\mathfrak{su}(2)\times{}\mathfrak{u}(1)$ subalgebra of $\mathfrak{su}(4)$. Operators belonging to it map a given X state into another one, and the above-given Kraus operators of the amplitude damping channel \eqref{eq:Kraus_AD-2qubit} belong to such subalgebra.

%%%%%%%%%%%%%%%%%%%%%%%%%%%%%%%%%%%%%%%%%%%%%%
%%%%%%%%%%%%%%%%%%%%%%%%%%%%%%%%%%%%%%%%%%%%%%
%%%%%%%%%%%%% Werner States %%%%%%%%%%%%%%%%%%%
%%%%%%%%%%%%%%%%%%%%%%%%%%%%%%%%%%%%%%%%%%%%%%
%%%%%%%%%%%%%%%%%%%%%%%%%%%%%%%%%%%%%%%%%%%%%%
\subsection{Werner states}\label{sec:Werner_AD}

Werner states, $\rho_w$, can be written as:
\begin{equation}\label{eq:rhoWerner}
\rho_w= z\dyad{\Psi^-} +\frac{1-z}{4}\,\mathbbm{1}_4,
\end{equation}
\noindent{}where $z\in[0,1]$ and $\ket{\Psi^-}=\frac{1}{\sqrt{2}}\qty(\ket{01}-\ket{10})$ is the singlet, one of the four maximally entangled 2-qubit states known as Bell states. In previous references \cite{LAQC_BD, LAQC_Xstates-sym}, we analyzed the behavior of LAQC for such states under Markovian decoherence, considering global Depolarizing, Phase Damping, and Amplitude Damping channels. In each case we considered the same type of quantum operation acting locally on both subsystems with the same interaction parameter.

The state $\rho_w^{(AD)}$ resulting from applying the above Kraus operators via \eqref{eq:Kraus_Interaccion} is a non-symmetric X state with non-null Bloch parameters given by
\begin{equation}\label{eq:WernerAD-Bloch}
\rho_w^{(AD)}:\;\;\begin{cases}
y_3 = p,\\
T_1 = T_2 = -\sqrt{1-p\,}\,z, \\
T_3 = -\qty(1-p)\,z.
\end{cases}
\end{equation}

From eq. \eqref{eq:LAQC-Asymm-Xstates}, it is direct to verify that
\begin{align}
        \mathcal{L}\qty(\rho_w^{(AD)}) =\,&\frac{1+\sqrt{1-p\,}\,z}{2}\log_2\qty[1+\sqrt{1-p\,}\,z]\\ &\,+\frac{1-\sqrt{1-p\,}\,z}{2}\log_2\qty[1-\sqrt{1-p\,}\,z].\nonumber
\end{align}
Using \eqref{eq:Concurrencia_X}, the concurrence is given by
\begin{equation}
    \mathcal{C}_{w^{AD}} = \max\qty{0,C_1},
\end{equation}
where
\begin{align*}
    C_1=\,&\frac{1}{4}\bigg(2z\sqrt{1-p\,}\\
    &\;\;\;- \sqrt{(1-p)(1-z)\qty\big[1+p-z(1-p)]\,}\bigg),
\end{align*}
and its QD \eqref{eq:QD-Def-B}, computed using the algorithm proposed by Quesada et al. \cite{Quesada-XStates}, is given by
\begin{equation}
\begin{aligned}
    D_B\qty(\rho_w^{(AD)}) =\,&\frac{3(1-z)}{4}\log_2\qty[\frac{1-z}{4}]\\
    &\;+\frac{1+3z}{4}\log_2\qty[\frac{1+3z}{4}]\\
    &\;-\frac{1-p}{2}\log_2\qty[1-p]\\ &\;-\frac{1+p}{2}\log_2\qty[1+p] +N_1,
\end{aligned}
\end{equation}
where
\begin{align*}
    N_1 =\,& 1 -\frac{1}{2}\log_2\qty[(p-1) z^2+1]\\ &\qq{ }+\frac{z\sqrt{1-p\,}}{\ln(2)}\, \mathrm{arctanh}\qty(z\sqrt{1-p\,}).
\end{align*}

\begin{figure}[ht]
\centering
\includegraphics[width=0.3\textwidth]{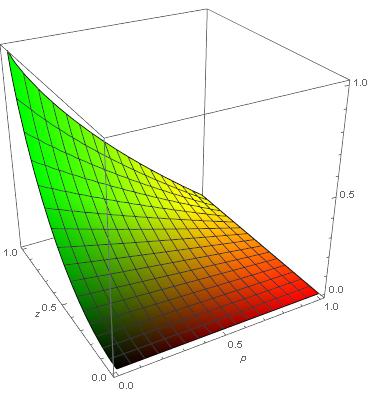}\\
\includegraphics[width=0.3\textwidth]{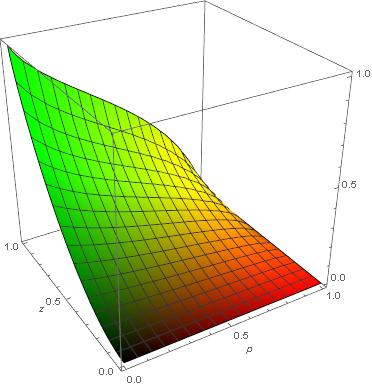}\\
\includegraphics[width=0.3\textwidth]{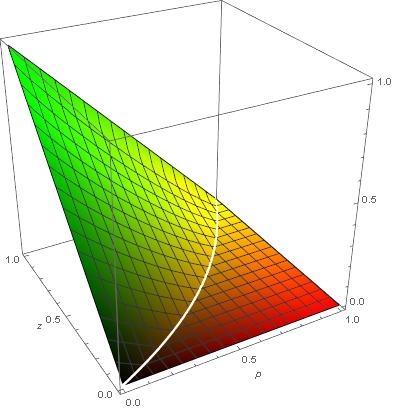}
\caption{LAQC (above), QD (middle), and Concurrence (below) for a Werner state under local amplitude damping.\label{fig:Werner-ADChannel}}
\end{figure}

In Figure \ref{fig:Werner-ADChannel} we present the graphical behavior of LAQC, QD, and concurrence. As was the case with other quantum channels, in particular global AD, Werner states again exhibits `Entanglement Sudden Death' \cite{EntSuddenDeath}. For LAQC and QD, these quantum correlations only vanish asymptotically. 

\begin{figure}[ht]
\centering
\includegraphics[width=0.35\textwidth]{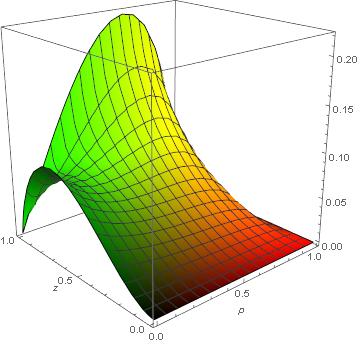}
\caption{Surface $\mathcal{S}(z,p)$ \eqref{eq:S(z,p)} resulting from the difference between QD and LAQC for a Werner state under local amplitude damping.\label{fig:Werner-ADChannel-QD-LAQC}}
\end{figure}

As was observed in other cases, QD is larger than LAQC, as can be seen in Figure \ref{fig:Werner-ADChannel-QD-LAQC}, where we show the surface 
\begin{equation}\label{eq:S(z,p)}
    \mathcal{S}(z,p)=D_A\qty(\rho_w^{(AD)}) -\mathcal{L}\qty(\rho_w^{(AD)}).
\end{equation}
We have that $\mathcal{S}(z,p)\geq0$ for all $z\in[0,1]$ and $p\in[0,1]$.

%%%%%%%%%%%%%%%%%%%%%%%%%%%%%%%%%%%%%%%%%%%%%%
%%%%%%%%%%%%%%%%%%%%%%%%%%%%%%%%%%%%%%%%%%%%%%
%%%%%%%%%%%%% Gen X states %%%%%%%%%%%%%%%%%%%
%%%%%%%%%%%%%%%%%%%%%%%%%%%%%%%%%%%%%%%%%%%%%%
%%%%%%%%%%%%%%%%%%%%%%%%%%%%%%%%%%%%%%%%%%%%%%
\subsection{General X states}

In \cite{Hu-LocalCreation}, Hu et al. derived the necessary and sufficient conditions for the local creation of quantum correlations. The amplitude damping channel is among the quantum operations that satisfies those conditions, and it has been established that it can create quantum discord \cite{Abad_2012-QD_Power, Xu_2012-QD_ADChannel, QD_creation-AD_2013}. Whether LAQC can be locally created or not is still an open question and our present results allows us to address this questions for the amplitude damping channel.

The state $\rho'$ resulting from applying this quantum operation to a general X state with Bloch parameters $\qty{x_3,y_3,T_1,T_2,T_3}$ via \eqref{eq:Kraus_AD-2qubit} has the following non-null Bloch parameters:
\begin{equation}\label{eq:Xstate_AD-Bloch}
\begin{aligned}
    &x_3' = x_3, &T_1'&=\sqrt{1-p\,} \,T_1,\\
    &y_3'= (1-p)y_3 +p\qc&T_2'&=\sqrt{1-p\,} \,T_2,\\
    &&T_3'&=(1-p)T_3 +px_3.
\end{aligned}
\end{equation}

Since the LAQC quantifier \eqref{eq:LAQC-Asymm-Xstates} of $\rho'$ is a function of $T_2'$ only, then it is straightforward to verify
\begin{equation}\label{eq:rhoX_AD-inequality}
    \mathcal{L}\qty(\rho^{X})\geq\mathcal{L}\qty\big(\rho') \qq{for all} p\in[0,1],
\end{equation}
whether $\rho^{X}$ is symmetric, antisymetric, or non-symmetric.

If the initial state is non-symmetric, the above inequality is trivial to verify. On the other hand, if $\rho^{X}$ is symmetric or anti-symmetric, we have to resort to the LAQC quantifier derived in \cite{LAQC_Xstates-sym}:
\begin{equation}\label{eq:LAQC-EstadosX-s_as}
    \mathcal{L}\qty(\rho^{X_{\smqty{s\\ as}}}) = \max\qty(g_j,g_{\pm}),
\end{equation}
where
\begin{subequations}\label{eq:g1g2g+}
\begin{align}
   g_j\equiv\,&  \frac{1+T_j}{2}\log_2\qty(1+T_j) + \frac{1-T_j}{2}\log_2\qty(1-T_j),\label{eq:gi(Ti)}\\
   g_{\pm}\equiv\,& \frac{1\pm{}T_3 + 2 x_3}{4}\log_2\qty[\frac{1\pm{}T_3 + 2 x_3}{(1+x_3)^2}]\nonumber\\
    &\;\;+\frac{1\pm{}T_3 - 2 x_3}{4}\log_2\qty[\frac{1\pm{}T_3 - 2 x_3}{(1-x_3)^2}]\nonumber\\ 
    &\;\;+ \frac{1-T_3}{2}\log_2\qty(\frac{1-T_3}{1-x_3^2})\label{eq:g+},
\end{align}
\end{subequations}
for $j=1,2$, and with $g_{+}$ for the symmetric case and $g_{-}$ for the anti-symmetric one. From eq. \eqref{eq:LAQC-EstadosX-s_as}, we have that
\begin{equation}
    \mathcal{L}\qty(\rho^{X_{\smqty{s\\ as}}})\geq{}g_2\geq{}g_2',
\end{equation}
where $g_2'$ refers to the function \eqref{eq:gi(Ti)} with $T_2'$ \eqref{eq:Xstate_AD-Bloch}. Therefore, it is straightforward to realize that eq. \eqref{eq:rhoX_AD-inequality} holds. Thus, the action of a local amplitude damping channel cannot create LAQC.

%%%%%%%%%%%%%%%%%%%%%%%%%%%%%%%%%%%%%%%%%%%%%%
%%%%%%%%%%%%%%%%%%%%%%%%%%%%%%%%%%%%%%%%%%%%%%
%%%%%%%%%%%%% Conclusiones %%%%%%%%%%%%%%%%%%%
%%%%%%%%%%%%%%%%%%%%%%%%%%%%%%%%%%%%%%%%%%%%%%
%%%%%%%%%%%%%%%%%%%%%%%%%%%%%%%%%%%%%%%%%%%%%%
\section{Conclusions}
We studied local available quantum correlations (LAQC) \cite{LAQC} for 2-qubit non-symmetric X states. In doing so, we obtained an exact and unique analytical expression of its quantifier. Alongside our previous results for symmetric and anti-symmetric X states \cite{LAQC_Xstates-sym}, the present results complete the study of the LAQC quantifier for 2-qubit X states.

We included the analysis of the action of a local amplitude damping channel. It has been established that such a channel can create quantum discord, but this is not the case for LAQC. This impossibility hints toward its monotonicity under LOCC operations. Therefore, formally demonstrating such monotonicity would be the next step in our research efforts. Also, the study of LAQC for qudit-qudit systems, i. e. qubit-qutrit and qutrit-qutrit states, as well as a multipartite version of LAQC for $N$-qubit systems ($N>2$) are still pending.

\section*{Acknowledgments}
This work was partially funded by the \emph{2020 BrainGain Venezuela} grant awarded to H. Albrecht by the \emph{Physics without Frontiers} program of the ICTP. The authors are thankful to Gloria Buendia for her comments and discussions. Albrecht and Bellorin would also like to thank the support given by the research group GID-30, \emph{Teoría de Campos y Óptica Cuántica}, at the Universidad Simón Bolívar, Venezuela.

\appendix*
\section{Quantum Discord of 2-qubit X states}

In \cite{Lu-QD-Xstates-Ex2} Lu et al. introduced the so-called maximal-correlation-direction measurements (MCDM) which were later used by Quesada et al. \cite{Quesada-XStates} to develop his algorithm to approximate the quantum discord of X states as 
\begin{equation}
    D_B\qty(\rho^X) \approx S\qty(\rho_B) - S\qty(\rho^X) + \min\qty{N_1, N_2},
\end{equation}
where $N_1$ \eqref{eq:N1-QD-Quesada} and $N_2$ \eqref{eq:N2-QD-Quesada} are the optimization functions derived using the MCDM. In this appendix we briefly summarize their procedure and how to determine these functions.

The starting point is to introduce a generic computational basis as the one given in \eqref{eq:BaseOrtonormalGen} for subsystem $\vb{B}$ and defining the corresponding projectors
\begin{equation}
    \Pi^{(2)}_i \equiv \dyad{\mu_i^{(2)}}\qc \Pi^{(\vb{AB})}_i=\mathbbm{1}^{(1)}_2\otimes\Pi^{(2)}_i.
\end{equation}
The state $\rho^{X}$ is then rewritten as
\begin{equation}
\begin{aligned}
    \rho^X_{A|\Pi^B_i} &= \frac{1}{p_i}\, \Pi^{(\vb{AB})}_i\rho^X \Pi^{(\vb{AB})}_i,\\ 
    p_i &= \Tr[\Pi^{(\vb{AB})}_i\rho^X \Pi^{(\vb{AB})}_i],
\end{aligned}
\end{equation}
and then the conditional entropy
\begin{equation}\label{CondEntr}
    S\qty(A|\qty{\Pi^B_i}) = \sum_i p_i S\qty(\rho^X_{A|\Pi^B_i})
\end{equation}
is determined. As is done in \cite{Liao_QD}, the critical points of this equation that are independent of the density matrix's elements are studied. With them, the angles $\theta_2$ and $\phi_2$ corresponding to the minima are determined and the following optimization functions are found:

\begin{subequations}
\begin{align}
    N_1 =& 1 -\frac{1+\sqrt{x_3^2+T_1^2\,}}{2} \log_2 \qty(1+\sqrt{x_3^2+T_1^2\,})\nonumber\\
    &-\frac{1-\sqrt{x_3^2+T_1^2\,}}{2}\log_2 \qty(1-\sqrt{x_3^2+T_1^2\,}),\label{eq:N1-QD-Quesada}\\
    N_2 =\,& - a \log_2 \qty[\frac{a}{a+c}] - b \log_2 \qty[\frac{b}{b+d}] \nonumber\\
    &\;\;- c \log_2 \qty[\frac{c}{a+c}] - d \log_2 \qty[\frac{d}{b+d}],\label{eq:N2-QD-Quesada}
\end{align}
\end{subequations}
where $a,b,c$, and $d$ \eqref{eq:Estados_X-abcdzw} are given in terms of $x_3$, $y_3$, and $T_3$ by
\begin{subequations}
\begin{align}
    a & = \frac{1}{4}\qty(1+x_3+y_3+T_3),\\
    b& = \frac{1}{4}\qty(1+x_3-y_3-T_3),\\
    c& = \frac{1}{4}\qty(1-x_3+y_3-T_3),\\
    d& = \frac{1}{4}\qty(1-x_3-y_3+T_3).
\end{align}
\end{subequations}

\bibliographystyle{apsrev}
\bibliography{biblio-qit}
\include{biblio-qit}

\end{document}